\documentclass{article}

\usepackage{arxiv}

\usepackage[utf8]{inputenc} 
\usepackage{textcomp}
\usepackage{txfonts}
\usepackage{hyperref}       
\usepackage{url}            
\usepackage{booktabs}       
\usepackage{amsfonts}       
\usepackage{nicefrac}       
\usepackage{microtype}      
\usepackage{lipsum}
\usepackage{graphicx}
\usepackage{cite} 
\usepackage{xcolor}

\newcommand{\beginsupplement}{%
        \setcounter{table}{0}
        \renewcommand{\thetable}{S\arabic{table}}%
        \setcounter{figure}{0}
        \renewcommand{\thefigure}{S\arabic{figure}}%
     }

\title{High academic performance is associated with shorter sleep and later bedtimes for young adults}

\author{
  Sofia Dokuka \\
  Institute of Education\\
  National Research University\\ Higher School of Economics\\
  Moscow 101000, Russia \\
  \texttt{sdokuka@hse.ru} 
   \And
  Ivan Smirnov \\
  Institute of Education\\
  National Research University\\ Higher School of Economics\\
  Moscow 101000, Russia \\
  \texttt{ibsmirnov@hse.ru} 
}

\begin{document}
\maketitle

\begin{abstract}
Shorter sleep is known to be negatively associated with academic performance. However, this result has mostly been found in homogenous samples (e.g., students from one university) or when using relative measures of academic performance, such as grade point average. Consequently, the relationship between academic performance and sleep patterns at the population level is not well understood. In this paper, we examined the relationship between academic performance as measured by a standardized test and sleep patterns using data from a Russian panel study ($N = 4,400$) that was nationally representative for one age cohort (20–21 years old). In addition to self-reported sleep patterns, the data set contained information about participants’ online activities over a period of up to 10 years, which allowed us to track the evolution of this relationship over time. We found that high academic performance was associated with shorter sleep, later bedtime, and increased online activity at night. The relationship between high academic performance and online activity at night was stable over a period of 5 years. Our findings suggest that the relationship between academic performance and sleep patterns can be more complex than previously believed and that high performance may be achieved at the expense of individual well-being.
\end{abstract}


\section*{Introduction}
In recent decades, an increasing proportion of the population has begun to experience sleep problems \cite{ford2015trends,jones2013us,de2016changes,tarokh2016sleep,keyes2015great,amra2017association} due to a combination of social \cite{smarr20183,hale2015recent,wheaton2016school,doane2015multi} and technological factors \cite{falbe2015sleep,vijakkhana2015evening,chassiakos2016children,cheung2017daily}. As poor sleep quality can negatively affect physical \cite{cappuccio2011sleep} and mental \cite{mayers2009subjective} health, as well as cognitive performance \cite{pilcher1996effects}, it is essential to understand the factors associated with poor sleep patterns. One particular area of interest is the association between sleep patterns and academic performance, as the latter is closely related to many important life outcomes \cite{organisation2013pisa,arendt2005does,gottfredson2004intelligence,roth1996meta,olsson201332}.

Academic performance may be affected by a mismatch between an individual’s natural diurnal patterns and socially constructed school schedules \cite{smarr20183,goldin2020interplay}. While humans significantly differ in terms of their preferred times for sleep and activity (known as their “chronotype”) \cite{wittmann2006social,adan2012circadian}, classes in primary and secondary schools and universities generally start early in the morning. Such discrepancies between chronotype and school schedule could result in poorer performance among individuals who prefer to be active in the evenings (so-called “social jet lag”) \cite{shochat2014functional,goldin2020interplay,wittmann2006social}. For instance, research has found that students with a sleep deficit receive lower grades \cite{short2013impact,preckel2011chronotype,hysing2016sleep} and that delaying school start times extends sleep duration and improves the academic performance of students with evening chronotypes \cite{wheaton2016school,gariepy2017school,goldin2020interplay}.

However, there is contradictory evidence regarding the relationship between chronotype and academic performance. Higher academic performance was attributed to both morning \cite{preckel2011chronotype,short2013impact,arbabi2015influence} and evening chronotypes \cite{roberts1999morningness,kanazawa2009night,panev2017association,piffer2014morningness}. To date, these studies have mostly focused on homogenous samples, i.e., students from one school or university \cite{goldin2020interplay,van2015timing}. Moreover, when studies are conducted on a larger scale, academic performance is usually measured in terms of grade point average (GPA) — for instance, in population-based studies of Norwegian adolescents (16–19 years old; $N = 7,798$) \cite{hysing2016sleep} and Finnish adolescents (11–18 years old; $N = 1,136,583$) \cite{kronholm2015trends}. As GPA is a relative measure of academic performance (that is, it is not standardized across different educational organizations), it does not allow for direct comparison of students from different schools or universities. Thus, the direction and strength of the relationship between academic performance and sleep patterns on the population level are not well understood.

In this paper, we used data from the Russian Longitudinal Panel Study of Educational and Occupational Trajectories (TrEC) \cite{malik2019russian}, which tracks 4,400 students who participated in the Programme for International Student Assessment (PISA) \cite{organisation2013pisa} in 2012. The TrEC is a nationally representative sample for one age cohort, who were 14–15 years old in 2012. In addition to academic performance, as measured by PISA scores, this data set includes questions related to participants’ sleep patterns. In 2018, participants were asked the following questions, based on the Munich Chronotype Questionnaire (MCTQ) \cite{roenneberg2003life}: “When do you usually go to bed on weekdays/on weekends?” and “When do you usually wake up on weekdays/on weekends?” In addition to self-reports, the data set contains digital traces (i.e., data from social media and other digital platforms) — specifically, information about participants’ online activity on the most popular Russian social networking site, VK, for those who agreed to share this data for research purposes (79\%). These digital traces are available for a period of up to 10 years and allow researchers to study the dynamics of the relationship between academic performance and sleep patterns over time.

We studied differences in sleep duration and bedtime among participants belonging to different academic proficiency levels, as defined by the Organisation for Economic Co-operation and Development (OECD) \cite{organisation2014pisa}. Since PISA scores were collected much earlier than information about sleep patterns, we were unable to estimate the influence of participants’ sleep patterns on their test results. Instead, we used PISA scores as an indicator of educational achievement—a characteristic that is known to be stable over time \cite{rimfeld2018stability} — and compared the sleep behavior of participants at different levels of academic proficiency. In our analysis, we also controlled for participants’ current occupational status (i.e., whether they were studying at university or working full-time) and socioeconomic status. We then used the times of participants’ most recent social media posts as a predictor of bedtime to study the evolution of the observed relationship over the previous 10-year period.

\section*{Results}
\subsection*{Self-reported sleep patterns and academic performance}
On average, participants went to bed at 11:43 p.m. and woke up at 7:41 a.m. on weekdays. Sleep patterns and academic proficiency were correlated (Fig. 1). The highest-performing men went to sleep 1 hour and 14 minutes later than the lowest-performing men, and the highest-performing women went to sleep 58 minutes later than the lowest-performing women. Higher-performing men appear to partially compensate for this later bedtime with a later wake time (1 hour and 3 minutes later than lower-performing men), but the same does not hold true for higher-performing women, who slept 42 minutes less than lower-performing women.

This result holds after we control for gender, socioeconomic status, and participants’ current occupational status in multiple regression models. Our models indicate that an 100-point increase in PISA score\footnote{PISA scores are normalized such that the mean is 500 and the standard deviation is 100 for OECD countries} is associated with a bedtime  that is approximately 20 minutes later (95\% CI [14.55, 21.90]; Fig. 2a) and a 10-minute reduction in sleep duration (95\% CI [5.74, 14.36]; Fig. 2b). Studying at university and working full-time decreased weekday sleep duration by 22 minutes (95\% CI [15.37, 29.00]) and 30 minutes (95\% CI [23.64, 36.52]), respectively. Men (47.1\%) went to bed 10 minutes later than women on average (95\% CI [4.40, 15.50]). Socioeconomic status did not have a statistically significant impact on bedtime or sleep duration.

Overall, we found that academic performance was negatively associated with sleep duration and positively associated with later bedtimes. Individuals with lower academic performance tended to go to bed and wake up earlier than higher-achieving individuals. High-achieving individuals do not appear to compensate for their late bedtime with later wake times. Studying in university and working full-time both had a negative impact on sleep duration.

The association between high academic performance and later bedtime may be explained by the fact that high-performing individuals are more active and have less free time. High-performing participants reported spending more time engaged in various activities during the day, including education, socializing, and entertainment. Spending greater amounts of time on these activities is known to be associated with later bedtimes \cite{ackerman2003time,knutson2009sociodemographic}. We found a similar relationship in our data set: Participation in both education and entertainment activities was positively associated with academic performance and bedtime (see SI).

\begin{figure}[h!]
\centering
\includegraphics[width=0.5\linewidth]{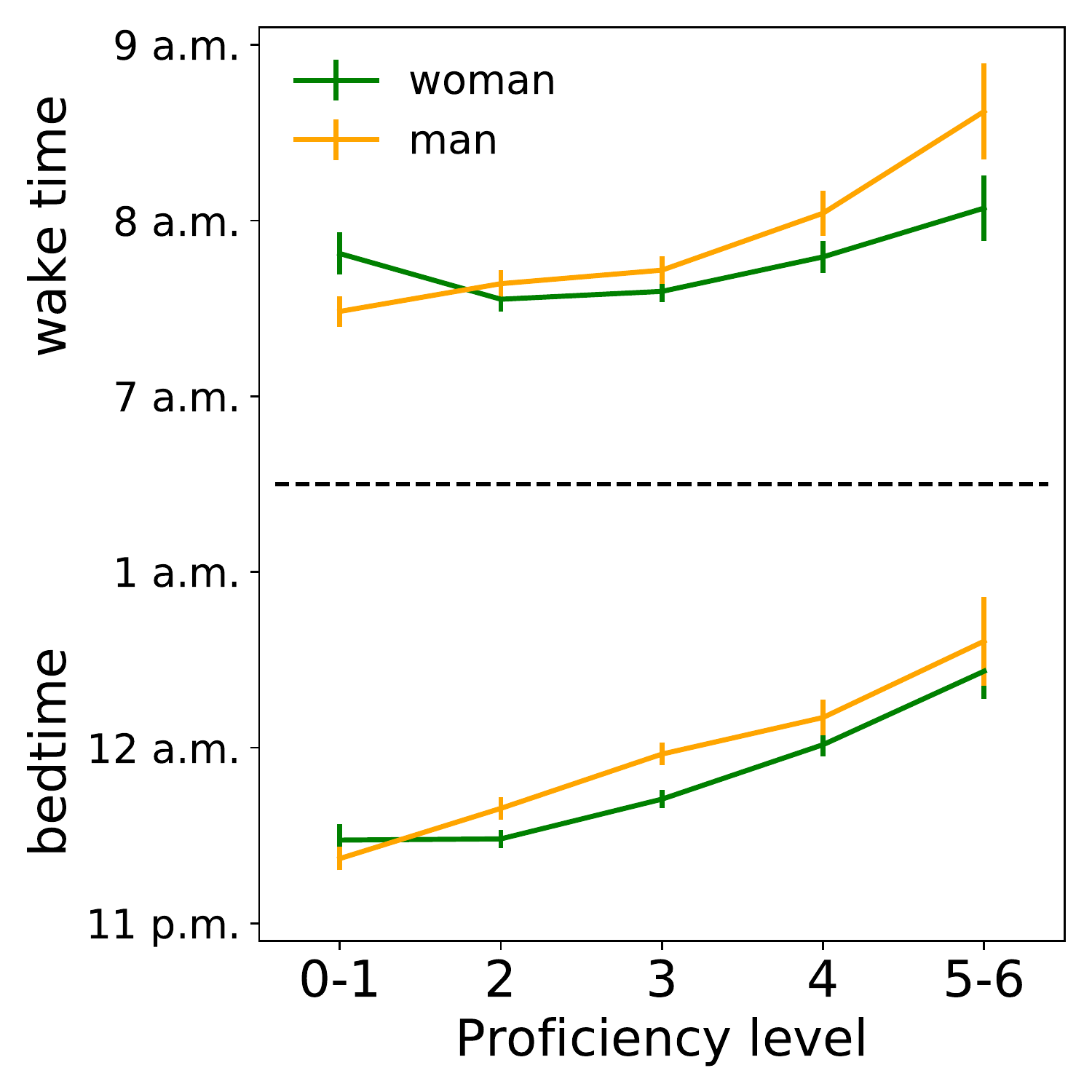}
\caption{\textbf{Average bedtime and wake times of survey participants as a function of proficiency level (as measured by PISA score).} Vertical bars represent standard errors. Academic performance and average bedtime are positively correlated: The highest-performing men go to bed 1 hour and 14 minutes later than the lowest-performing men, and the highest-performing women go to bed 58 minutes later than the lowest-performing women.}
\label{fig:levels}
\end{figure}

\begin{figure}[h!]
\centering
\includegraphics[width=1.1\linewidth]{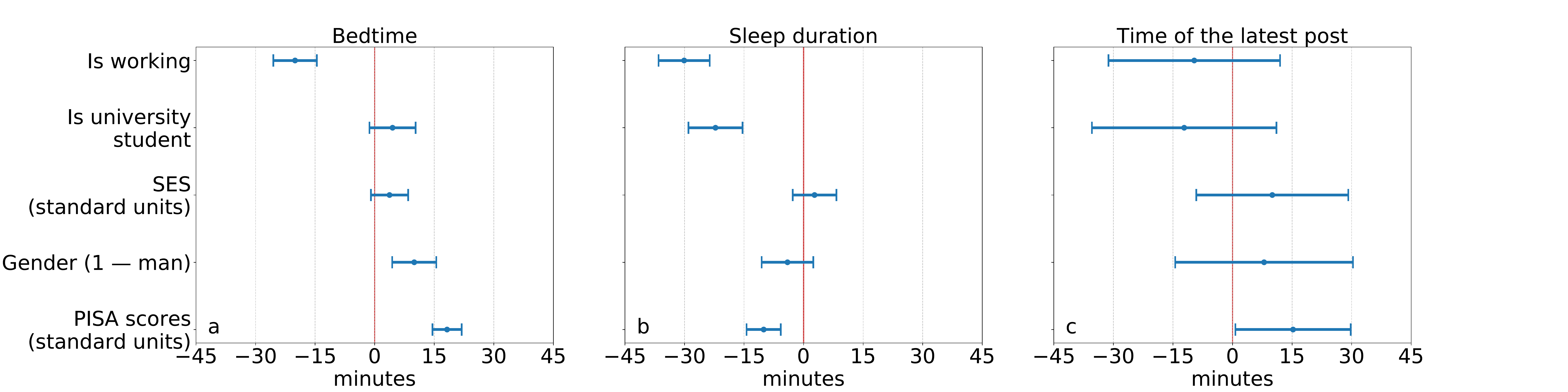}
\caption{\textbf{Effects of academic performance on sleep patterns.} Higher academic performance is associated with later bedtime, shorter sleep duration, and later time of most recent social media post. Estimates of coefficients along with 95\% confidence intervals from multiple regression models are shown for academic achievements and control variables.
a) Effects on bedtime (minutes after midnight; $N = 3,310$; intercept =  20.9 minutes).
b) Effects on sleep duration (minutes; $N = 3,310$; intercept = 506.5 minutes).
c) Effects on latest post time (minutes after midnight; $N = 514$; intercept = 48.7 minutes).
}
\label{fig:controls}
\end{figure}

\subsection*{Tracking the evolution of the relationship between bedtime and academic performance using digital traces}
We used participants’ digital traces to study the evolution of the relationship between bedtime and academic performance over time. Specifically, we collected the timestamps of public posts for participants who agreed to share their VK data ($N\textsubscript{users} = 3,483$; $N\textsubscript{posts} = 781,386$). The distribution of posting activity in a 24-hour period is presented in Figure 3. This distribution follows the expected patterns of human circadian rhythms \cite{liang2018birds,smarr20183}, with nearly half of all posts (49.4\%) published between 6:00 p.m. and 1:00 a.m. and only 8\% published between 2:00 a.m. and 9:00 a.m.

We found that the time of a participant’s most recent VK post during the survey data collection period was correlated with self-reported bedtime (Pearson’s $r = 0.23$; $P < 10^{-13}$; see Materials and Methods for more details). The relationship between time of latest post and academic performance was similar to the results obtained for self-reported bedtime: An 100-point increase in PISA score corresponded to 15 minutes later post time ($P = 0.04$, 95\% CI [0.73, 29.77]) after controlling for gender, socioeconomic status, and occupational status (Fig. 2c).

In addition, we compared the average time of latest post for higher- and lower-performing participants in each month over the last 10 years to reveal the nature of the relationship between chronotype and academic performance. According to our results, in the last 5 years, lower-performing students consistently wrote their latest posts on average 17 minutes earlier than higher-performing students ($P = 3.6 \times 10^{-5}$; Fig. 4). We did not find a similar difference for data from the first 5 years, where higher-performing students’ most recent post was just 1 minute later ($P = 0.89$). Using digital traces thus confirms the robustness of the observed tendency for high-achieving individuals to go to bed later over a prolonged period of time.

\begin{figure}[h!]
\centering
\includegraphics[width=0.5\linewidth]{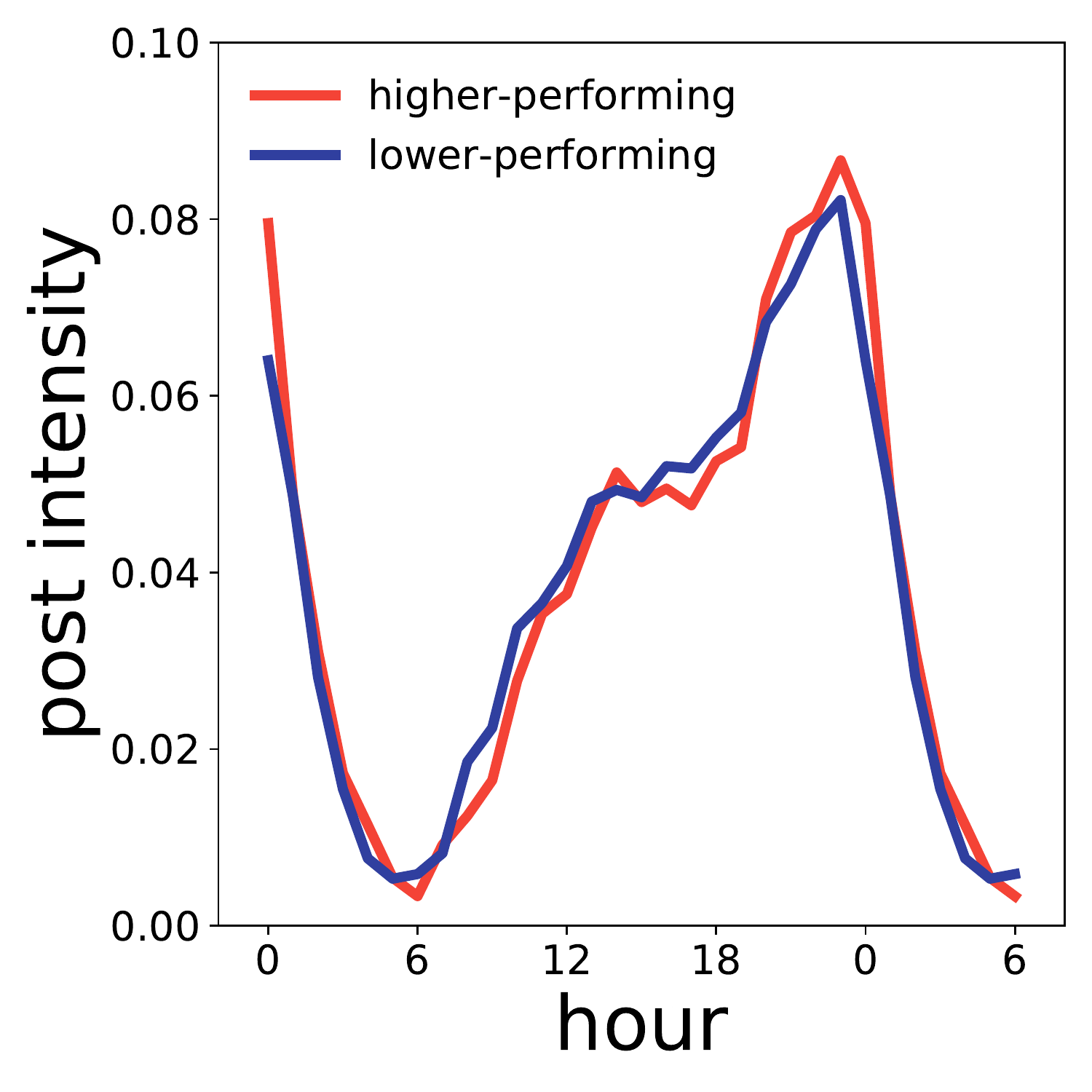}
\caption{\textbf{Changes in online activity during the day.} The average proportion of all social media posts that were written in a given hour of the day is shown for higher-performing (top half) and lower-performing (bottom half) students. Almost half of all posts (49.4\%) were published between 6:00 p.m. and 1:00 a.m., while only 8\% were published between 2:00 a.m. and 9:00 a.m. Higher-performing students wrote a larger proportion of their posts after midnight (between 12:00 a.m. and 5:00 a.m.) than lower-performing students: 15\% versus 13\% ($P = 0.015$). 
}
\label{fig:activity}
\end{figure}

\begin{figure}[h!]
\centering
\includegraphics[width=0.5\linewidth]{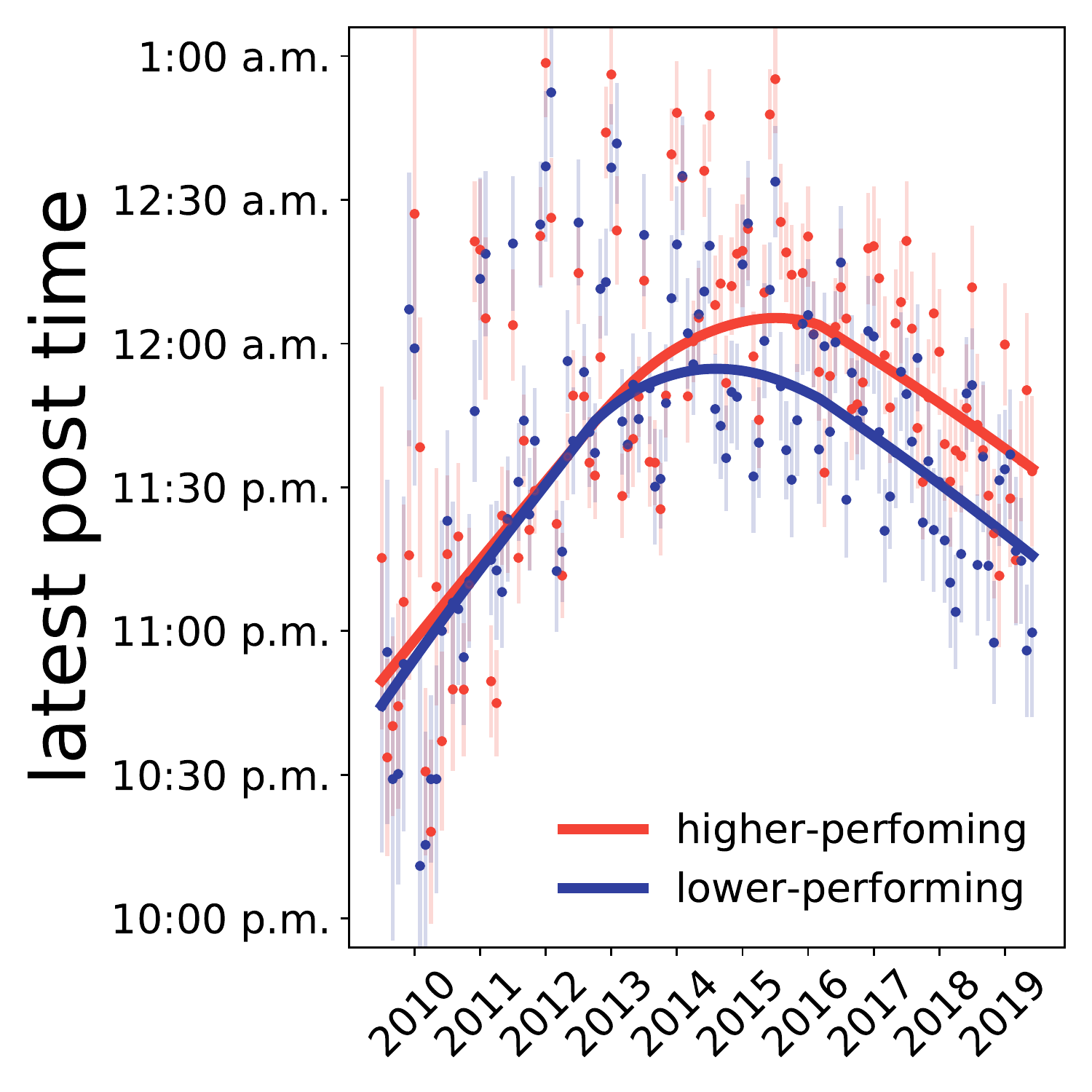}
\caption{\textbf{The evolution of latest social media post time for higher-performing (top half) and lower-performing (bottom half) participants.} Average values and standard errors were computed for each month for the past 10 years. The LOESS curve highlights the main trend. In the last 5 years, lower-performing students consistently wrote their most recent posts on average 17 minutes earlier ($P = 3.6 \times 10^{-5}$) than higher-performing students. In the first 5 years, higher-performing students’ most recent post was just 1 minute later ($P = 0.89$).
}
\label{fig:evolution}
\end{figure}

\section*{Discussion}
Individual sleep patterns have traditionally been studied using standardized questionnaires, such as the MCTQ and the Morningness-Eveningness Questionnaire (MEQ) \cite{roenneberg2003life,horne1976self}. These measurements are an easy and effective way to investigate sleep-related variables \cite{goldin2020interplay}. However, retrospective self-reports are limited in temporal granularity, might be affected by social desirability bias, and are vulnerable to memory error and experimenter demand effects \cite{smarr20183,lewis2015three,golder2011diurnal}.

In contrast, digital traces offer localized real-time observation of individual behavior \cite{golder2011diurnal}, thus providing actual information about people’s behavior in natural settings and enabling retrospective data collection. Overall, social media, online platforms, and metadata are becoming popular and rich data sources for detailed studies of individual diurnal patterns \cite{cuttone2017sensiblesleep,smarr20183}. Researchers can create specific proxy measures for sleep pattern identification, such as latest time of mobile phone use \cite{cuttone2017sensiblesleep} or the difference between intensity of online activity on school days versus non-school days \cite{smarr20183}. 

That said, digital traces also have some limitations. As data from digital platforms is not specifically collected for academic purposes, there is often no direct connection between recorded behavior and the theoretical construct being studied \cite{sen2019total,salganik2019bit}. For instance, sleep patterns are usually inferred from information about online activity \cite{liang2018birds,smarr20183}, ignoring the fact that some individuals might be active without being online.

In our study, we employed both survey and digital trace data to explore the relationship between academic performance and sleep patterns. This triangulation allowed us to confirm the robustness of the observed relationship. We found that higher-performing individuals slept less, tended to go to bed later, and were more active online at night than lower-performing individuals. This last correlation appears to have remained stable over the last 5 years. The observed relationship is partly explained by participation in various activities, both educational and entertaining in nature. High-performing individuals may attempt to include more activities in the day then lower-performing individuals, thus reducing the amount of time they have available for sleep.

As later bedtimes are known to be negatively associated with physical and mental health \cite{kuperczko2015late,merikanto2013late,gamble2013delayed,ohayon2005normative,mikulovic2014influence}, our results could indicate that high performance is achieved at the expense of individual health and well-being. Previous research has demonstrated that sleep can affect academic performance. Our findings suggest that this relationship might be more complex than previously believed and that academic performance may indirectly affect sleep patterns.

\section*{Materials and Methods}
\subsection*{Academic performance}
We used PISA reading scores as a measure of students’ academic performance. PISA defines reading literacy as “understanding, using, reflecting on and engaging with written texts in order to achieve one’s goals, to develop one’s knowledge and potential, and to participate in society” and considers it a foundation for achievement in other academic subjects, as well as a prerequisite for successful participation in most areas of adult life \cite{schleicher2009pisa}. PISA scores are scaled so that the OECD average is 500 with a standard deviation of 100, with every 40 points roughly corresponding to one year of formal schooling \cite{organisation2014pisa}. The PISA scale is divided into seven proficiency levels (0 to 6) \cite{organisation2014pisa}. Women outperformed men in reading in all studied countries. In our sample, women on average scored 40 points higher than men ($M_{women} = 497; M_{men} = 457; P < 10^{-46}$). As our sample included few women with a proficiency level of 0 and few men with a proficiency level of 6, we combined levels 0 and 1 and levels 5 and 6, thereby reducing the number of levels from seven to five (Fig. 1). When dividing participants into lower- and higher-performing groups (Figs. 3, 4), we normalized scores by gender to avoid confusing performance-related patterns with gendered patterns.

\subsection*{Digital traces collection}
We used VK’s API to download the timestamps of all public posts made by consenting participants between January 1, 2010, and December 31, 2019. Taking into account participants’ time zones, these timestamps were converted to the hour of the day when the posts were published with a resolution of 1 minute. To facilitate the comparison of digital traces with survey data, we considered only posts published on weekdays.

We computed the fraction of posts published at night and the time of the latest post published at night for the period of survey data collection (autumn 2018) and for each participant. For our purposes, “night” was defined as a period from T hours to 5:00 a.m. For various values of T, we computed the correlation between these two characteristics and self-reported bedtime (Fig. S1). While both variables had similar predicting power, the time of latest post appears to be less affected by the particular choice of T. This made the time of latest post a preferable proxy for bedtime for studying its evolution over a long time period. Changes in time of latest post are also easier to interpret than changes in the proportion of all posts written at night. For these reasons, we selected time of latest post at night as a proxy for bedtime and defined “night” as the period from 7:00 p.m. to 5:00 a.m.

\section*{Acknowledgements}
This work was supported by a grant from the Russian Science Foundation (project №19-18-00271).

The data of the Russian panel study “Trajectories in Education and Career” (TrEC http://trec.hse.ru/) is presented in this work. The TrEC project is supported by the Basic Research Programme of the National Research University Higher School of Economics.

\bibliographystyle{unsrt}
\bibliography{references}

\clearpage
\beginsupplement
\section*{Supplementary Information}
\subsection*{The relationship between activities, bedtime, and academic performance}
We used data about participants’ activities in 2017 and 2018 to test whether this information could explain the relationship between academic performance and bedtime. In 2017, participants were asked to estimate the amount of time they spent on different educational, social, and entertainment activities. The question was formulated as follows: “How many hours per week did you spend on the following activities in the 2016/2017 academic year? Scale from 1 to 8, where 1 corresponds to 0 hours, 8 corresponds to more than 30 hours.” In 2018, participants were also asked if they had “participate[d] in educational courses and workshops besides their studies in the last year.”
 
We found that spending time on hobbies, playing computer games, watching television and movies, social media usage, and online and self-education were positively associated with academic performance and bedtime (Table \ref{tab:activity}). Individuals who participated in educational activities (33.9\%) outperformed individuals who did not take part in such courses ($M_{activity} = 502$; $M_{no~activity} = 469$; $P < 10^{-28}$) and reported later bedtimes ($M_{activity} = $11:48 p.m.; $M_{no~activity} = $11:24 p.m.; $P < 10^{-6}$). These results held after we controlled for participants’ academic performance, gender, socioeconomic status, and current occupational status in multiple regression models (Fig. \ref{fig:activity_regr}). The model indicates that participation in educational courses is associated with a bedtime that is approximately 11 minutes later (95\% CI [ 4.79, 17.74]). If activities are included in the regression model, the regression coefficient corresponding to academic performance decreases from 18.2 to 13.8 minutes. This indicates that the relationship between academic performance and bedtime might be partially explained by participation in educational and entertainment activities.

\begin{figure}[h!]
\centering
\includegraphics[width=0.5\linewidth]{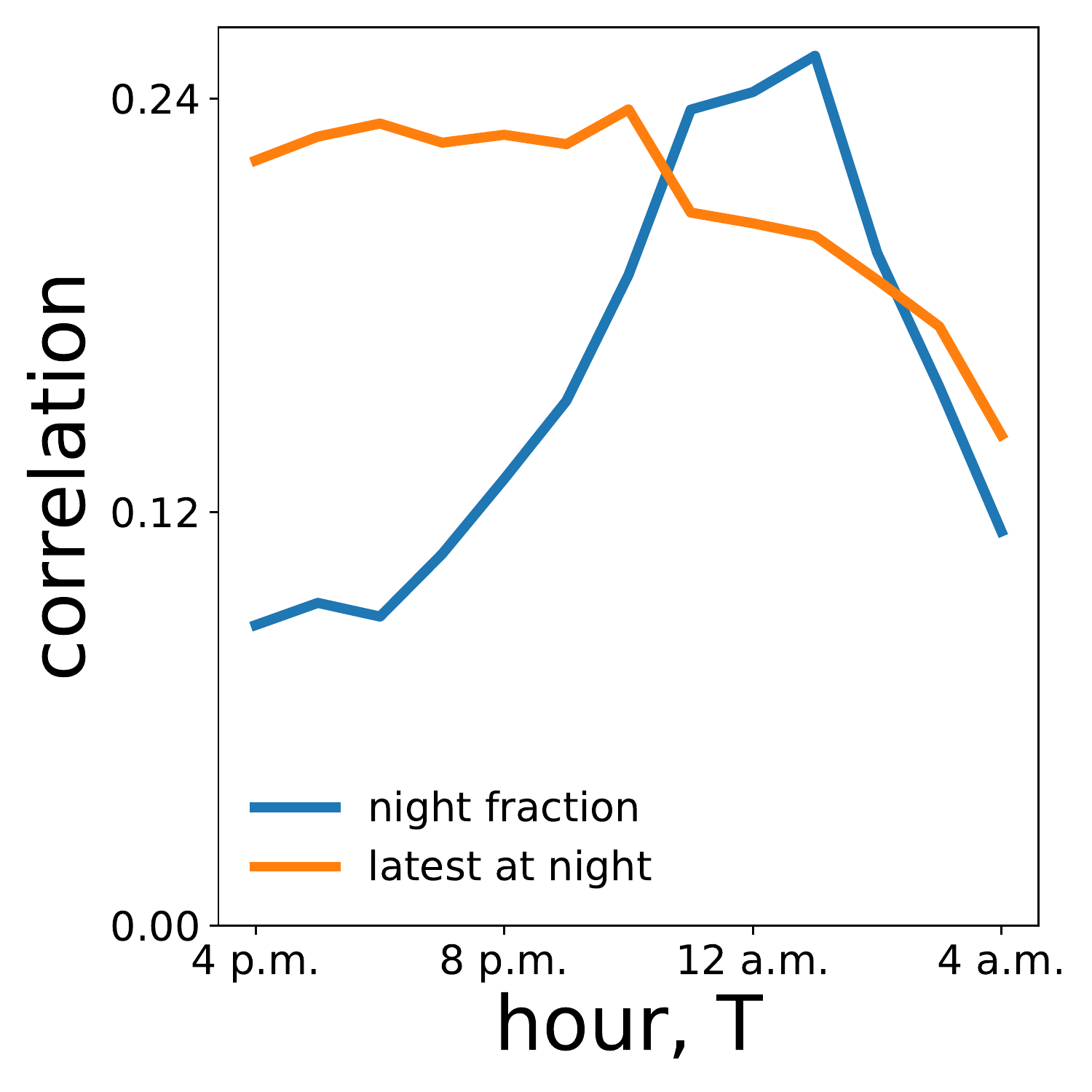}
\caption{This figure presents the relationships between self-reported bedtime and fraction of posts written at night (blue line) or the time of latest post at night (orange line). “Night” is defined here as a period from $T$ to 5:00 a.m. The results are computed for different values of $T$ (hour axis). Both variables have similar predicting power, but the time of latest post appears to be less affected by the particular choice of $T$.
}
\label{fig:measure}
\end{figure}

\begin{figure}[h!]
\centering
\includegraphics[width=0.8\linewidth]{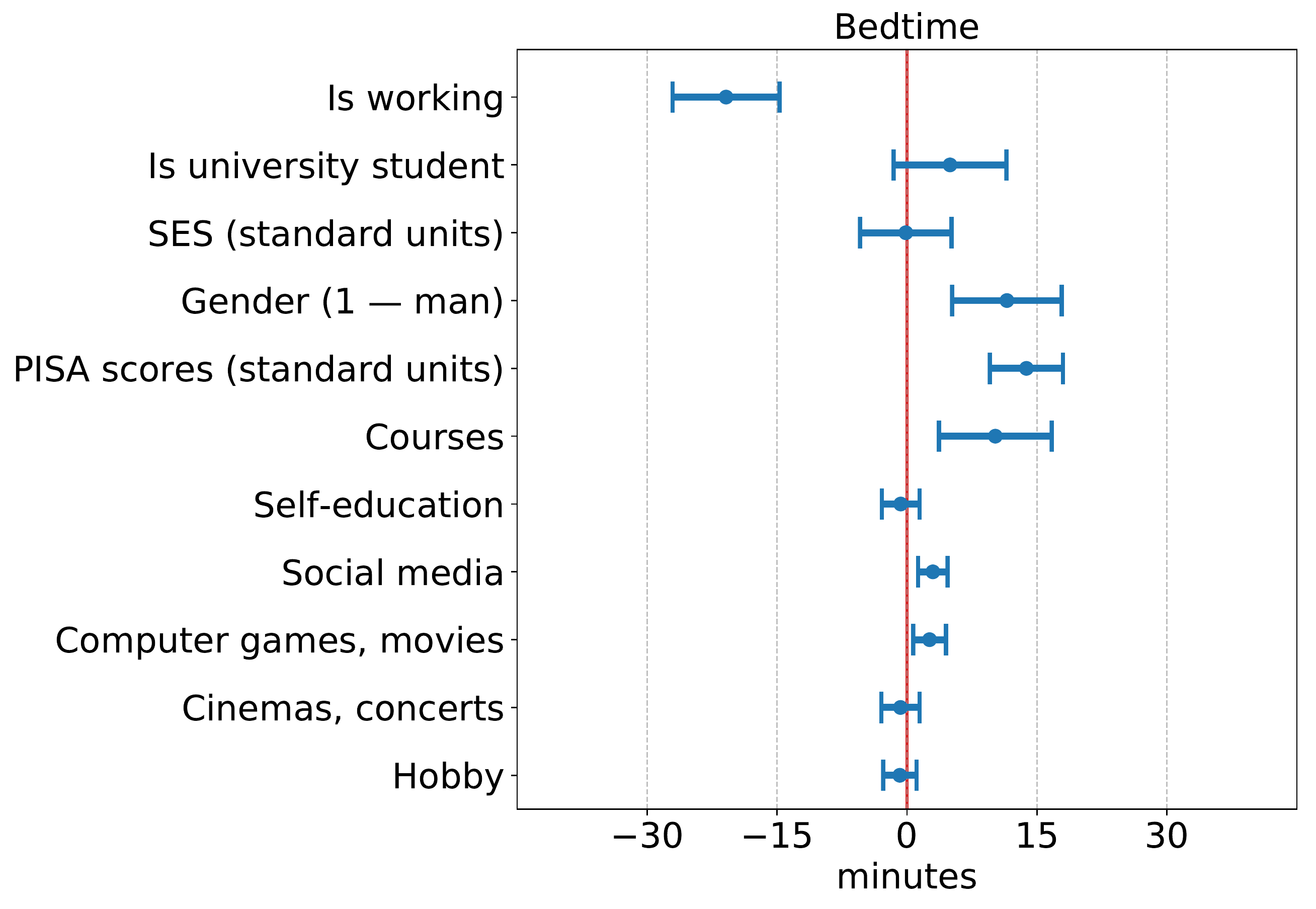}
\caption{Educational and entertainment activities are positively associated with academic performance and bedtime. Estimates of coefficients along with 95\% confidence intervals from multiple regression model are shown for activities, academic achievements and control variables. Effects on bed time on working days (minutes after midnight; $N = 2,565$; intercept = -42.0 minutes). }
\label{fig:activity_regr}
\end{figure}

\begin{table}
\centering
\caption{Pearson’s correlation coefficients between participation in various activities and academic performance, bedtime}
\begin{tabular}{lrr}
\textbf{Activity} & \textbf{Academic Performance} & \textbf{Bedtime}  \\
\hline \\
Hobbies (including creative courses)                    & $0.03$\textcolor{white}{$^{*}$} & $0.04$\textcolor{white}{$^{*}$}\\
Visiting cinemas, concerts, sports events, and other entertainment events &
-0.01\textcolor{white}{$^{*}$} & 0.02\textcolor{white}{$^{*}$}  \\
Playing computer games, watching movies and TV shows on the Internet                                               & $0.07^{*}$      & $0.11^{*}$           \\
Using social media                                         & $0.10^{*}$       & $0.11^{*}$           \\
Online education, self-education                           & $0.14^{*}$  & $0.05^{*}$
\\ \\
$^{*}$ -- $P < 0.01$
\end{tabular}
\label{tab:activity}
\end{table}

\end{document}